\documentclass[manuscript]{aastex}
\usepackage{times, graphicx}

\shorttitle{Source of propagating slow waves}
\shortauthors{Krishna Prasad et al.}

\begin{document}
\title{On the Source of Propagating Slow Magneto-acoustic Waves in Sunspots}             
\author{S. Krishna Prasad} 
\affil{Astrophysics Research Centre, School of Mathematics and Physics, Queen's University Belfast, Belfast, BT7 1NN, UK.}                                  
\email{krishna.prasad@qub.ac.uk}
\author{D. B. Jess} 
\affil{Astrophysics Research Centre, School of Mathematics and Physics, Queen's University Belfast, Belfast, BT7 1NN, UK.}                                  
\affil{Department of Physics and Astronomy, California State University Northridge, Northridge, CA 91330, U.S.A.}                                  
\author{Elena Khomenko} 
\affil{Instituto de Astrof{\'{i}}sica de Canarias, 38205 La Laguna, Tenerife, Spain.}                                  

\begin{abstract}
Recent high-resolution observations of sunspot oscillations using simultaneously operated ground- and space-based telescopes reveal the intrinsic connection between different layers of the solar atmosphere. However, it is not clear whether these oscillations are externally driven or generated in-situ. We address this question by using observations of propagating slow magneto-acoustic waves along a coronal fan loop system. In addition to the generally observed decreases in oscillation amplitudes with distance, the observed wave amplitudes are also found to be modulated with time, with similar variations observed throughout the propagation path of the wavetrain. Employing multi-wavelength and multi-instrument data we study the amplitude variations with time as the waves propagate through different layers of the solar atmosphere. By comparing the amplitude-modulation period in different layers, we find that slow magneto-acoustic waves observed in sunspots are externally driven by photospheric p-modes, which propagate upwards into the corona before becoming dissipated. 
\end{abstract}

\keywords{magnetohydrodynamics (MHD) --- Sun: atmosphere --- Sun: oscillations --- Sun: photosphere --- sunspots}

\section{Introduction}
Waves and oscillations are an integral part of sunspots. To date there have been many observational reports of oscillations in the photosphere, umbral flashes and running penumbral waves in the chromosphere, and propagating waves in the corona \citep[see, e.g., the review articles by][]{2006RSPTA.364..313B, 2015SSRv..tmp...14J, 2015lsrp_tmp}. Oscillations manifesting in photospheric sunspots usually display prominent 5~min periodicities alongside traces of power in 3~min bandpasses, albeit with relatively smaller amplitudes compared to surrounding quiet Sun \citep{1972SoPh...27...80B, 1984ssdp.conf..141A, 1997AN....318..129L, 2000ApJ...534..989B}. Umbral flashes \citep{1969SoPh....7..351B} are approximately 3~min periodic brightenings in chromospheric umbrae that occur as a result of upwardly propagating magneto-acoustic oscillations converting into localised shock waves \citep{2003A&A...403..277R, 2015A&A...574A.131H}. 

Running penumbral waves are outwardly propagating oscillations found in the chromospheric penumbrae of sunspots \citep{1972ApJ...178L..85Z, 1972SoPh...27...71G}. The periodicities of these waves, which are on the order of a few min, increase from the inner penumbra to the outer boundary, while the associated phase speeds decrease \citep{1997ApJ...478..814B, 2000A&A...354..305C, 2013ApJ...779..168J}. It is believed that the `trans-sunspot' (i.e., outward) motion is apparent to a given line-of-sight, and that these oscillations actually represent the upward propagation of field-guided magneto-acoustic waves from the photosphere \citep{2001A&A...375..617C, 2006SoPh..238..231K, 2007ApJ...671.1005B, 2013ApJ...779..168J}. The gradual change in inclination of the penumbral field lines produces the observed changes in the oscillation periods and phase speeds. Recently, \citet{2015arXiv150309106L} identified the photospheric signatures of running penumbral waves and found them to be consistent with the upward propagation of magneto-acoustic waves predicted previously. \citet{2012ApJ...746..119R} have even suggested frequency fluctuations within umbrae itself indicate substantial local variations in magnetic field inclinations.

Evidence for propagating waves along fan-like loop structures in the corona is in abundance. The fan loops associated with sunspots are usually rooted in the umbra, and display propagating waves with periodicities approximately equal to 3~min \citep{2002A&A...387L..13D, 2006ApJ...643..540M, 2012ApJ...746..119R,2012ApJ...757..160J}. The amplitude of these waves decreases rapidly as they propagate along the loop, eventually disappearing after several thousand km. Several physical processes such as thermal conduction and compressive viscosity are believed to dissipate such waves in the corona \citep{2003A&A...408..755D, 2014ApJ...789..118K}. \citet{2012ApJ...757..160J} found an association between 3~min waves propagating along fan loops and simultaneous amplitude enhancements in underlying photospheric umbral dots. Their results suggest that magnetic-acoustic waves manifesting in the photosphere can propagate upwards into the corona where, ultimately, the energy they carry is dissipated.

As advancements are being made in solar instrumentation and observations, it is becoming increasingly evident that all the above phenomena detected in different layers of the solar atmosphere are actually inter-connected and most likely produced by the same upwardly propagating slow magneto-acoustic waves, which present themselves according to the local physical conditions. Although it is often assumed that the photospheric $p$-modes are the ultimate source of sunspot oscillations, there is no clear evidence as to whether these oscillations are externally driven by the $p$-modes, or generated in-situ within the sunspots (e.g., through magneto-convection). In this Letter, we aim to address this issue by studying temporal variations in the amplitudes of propagating slow waves observed in a coronal fan loop which were not explored before. We present the details on our observations, our analysis methods and results, in the subsequent sections and finally discuss the conclusions.

\section{Observations}
The Dunn Solar Telescope (DST) at Sacramento Peak, New Mexico, was employed to observe active region NOAA~11366 on 2011 December 10 between 16:10 -- 17:25 UT. The Rapid Oscillations in the Solar Atmosphere \citep[ROSA;][]{2010SoPh..261..363J} and the Hydrogen-Alpha Rapid Dynamics camera \citep[HARDcam;][]{2012ApJ...757..160J} instruments were employed to simultaneously capture high-resolution images in four different optical channels centred at the H$\alpha$ line core (6562.8{\,}\AA), Ca \textsc{ii} K line core (3933.7{\,}\AA), G-band (4305.5{\,}\AA), and blue continuum (4170{\,}\AA) wavelengths, with filter bandpasses corresponding to 0.25{\,}\AA, 1{\,}\AA, 9.2{\,}\AA, and 52{\,}{\AA}, respectively. The pixel scale was $0{\,}.{\!\!}{\arcsec}0696$ for the ROSA channels, producing a field-of-view equal to $69\arcsec \times 69\arcsec$. For H$\alpha$ images acquired by HARDcam, the plate scale was $0{\,}.{\!\!}{\arcsec}138$ per pixel, providing a marginally larger field-of-view of $71\arcsec \times 71\arcsec$. Part of this dataset has been used previously by \citet{2013ApJ...779..168J}, where the authors detail the full speckle reconstruction and calibration steps applied to the data. Following all image processing steps, the final cadences of the data are 2.11~s for the 4170{\,}{\AA} continuum and G-band channels, 7.39~s for the Ca~\textsc{ii}~K filtergrams and 1.78~s for the narrowband H$\alpha$ image sequence. Seeing conditions remained excellent during the 75~min observation period. However, a few images were affected by local and short duration atmospheric fluctuations, which resulted in slight image degradation in locations away from the adaptive optics lock point. These images, corresponding to an average duration not longer than a few seconds, were replaced through interpolation. The co-alignment between the different ROSA channels is achieved using a series of collimated targets obtained immediately after the end of the science observations.

The corresponding space-based data from the Atmospheric Imaging Assembly \citep[AIA;][]{2012SoPh..275...17L} and the Helioseismic and Magnetic Imager \citep[HMI;][]{2012SoPh..275..229S}, onboard the Solar Dynamics Observatory \citep[SDO;][]{2012SoPh..275....3P}, forms the main part of the present study. Level 1.0 data from both instruments were processed using the routines \texttt{aia\_prep.pro} and \texttt{hmi\_prep.pro} available through standard solar software (e.g., {\sc{sswidl}}) pipelines. This involves bringing all the data to a common centre and plate scale, with a final pixel scale $\approx$$0{\,}.{\!\!}{\arcsec}6$. A subfield of $210\arcsec \times 210\arcsec$ is then carefully selected (by accounting for solar rotation) around the target region from AIA 171{\,}\AA, 131{\,}\AA, 304{\,}\AA, 1600{\,}\AA, and 1700{\,}{\AA} channels, in addition to HMI Dopplergrams covering the full observational duration. The cadence is 12~s for the AIA 171{\,}\AA, 131{\,}\AA, and 304{\,}{\AA} channels, 24~s for the AIA 1600{\,}{\AA} and 1700{\,}{\AA} channels, and 45~s for the HMI Dopplergram data, with each image co-aligned to the first image using intensity cross-correlation. HMI and ROSA (4170{\,}{\AA}) continuum images were used to achieve the required co-alignment between ground- and space-based data. We did not use the broadband 4170{\,}{\AA} continuum channel for subsequent analyses since the ROSA G-band data originates very close to the continuum level \citep[$\sim$75~km height difference;][]{2012ApJ...746..183J} and has a better signal-to-noise ratio than the ROSA 4170{\,}{\AA} continuum.

\section{Analysis and Results}
\begin{figure}
\centering
\includegraphics[scale=0.69]{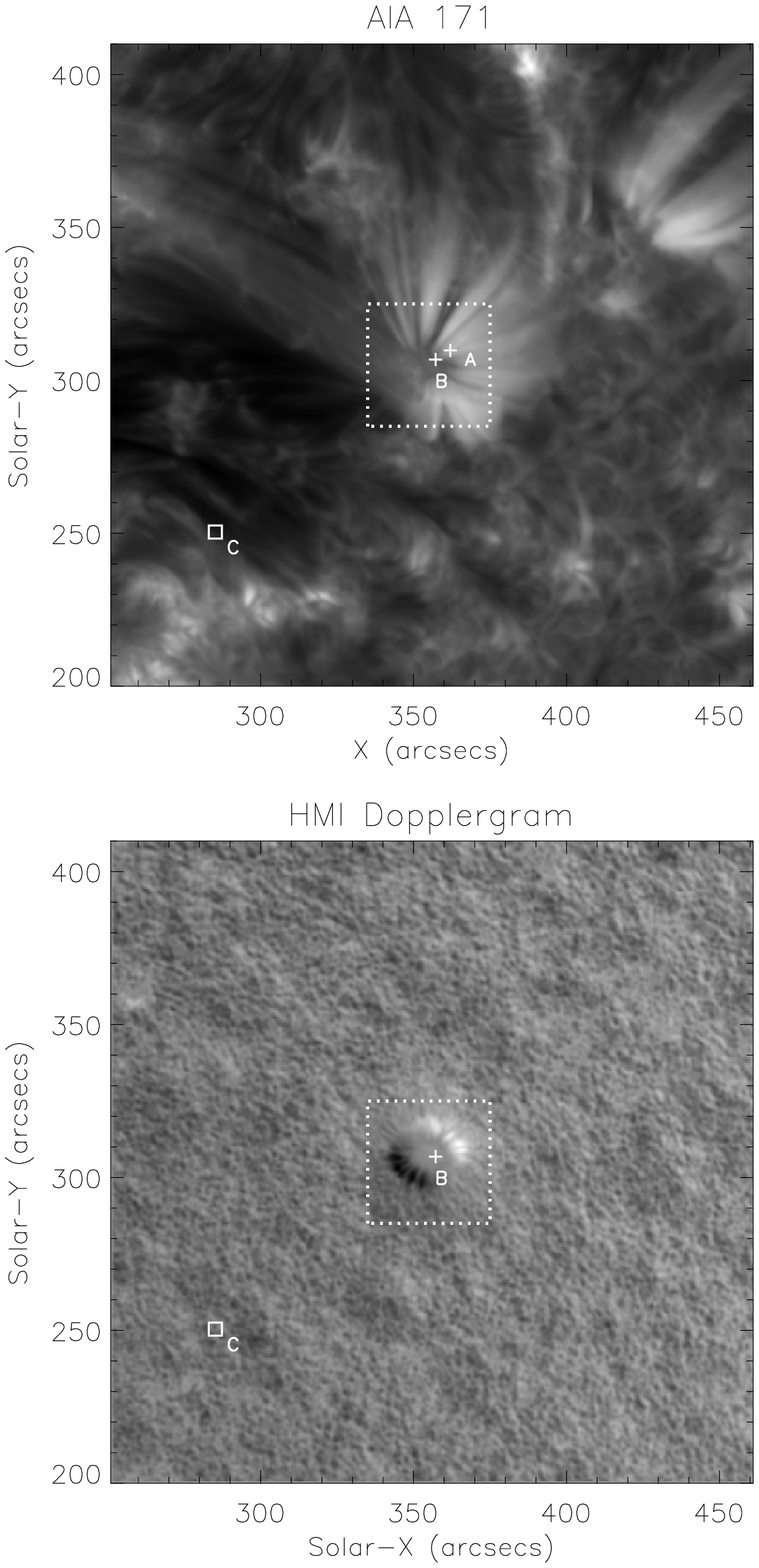}
\includegraphics[scale=0.51]{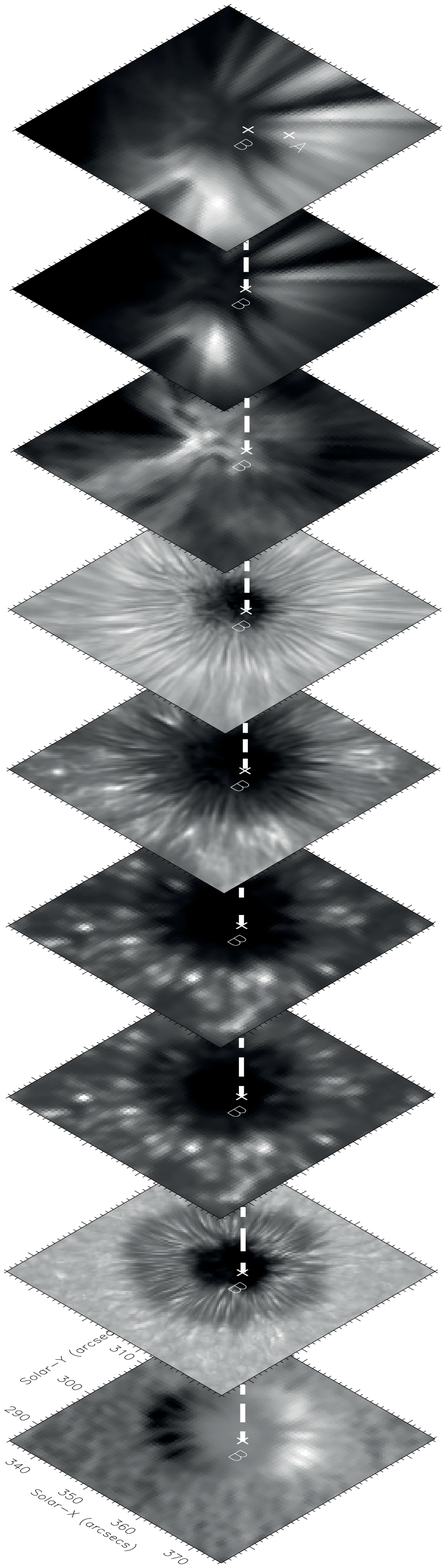}
\caption{\textit{Left:} Active region NOAA~11366 and its surroundings observed in the AIA~171{\,}{\AA} channel (top) and an HMI Dopplergram (bottom). The dotted box encloses the subfield region shown as a series of stacked images on the right. \textit{Right:} The sunspot as seen in different imaging channels: (from bottom) HMI Dopplergram, ROSA G-band, AIA~1700{\,}{\AA}, AIA~1600{\,}{\AA}, ROSA Ca~\textsc{ii}~K, HARDcam H$\alpha$, AIA~304{\,}{\AA}, AIA~131{\,}{\AA} and AIA~171{\,}{\AA}. Locations labelled `A', `B', and `C' are also marked, which are used to study wave propagation in subsequent Figures. A time-lapse movie of the active region in AIA 171{\,}{\AA} channel can be found online.}
\label{fig1}
\end{figure}
Active region NOAA~11366 comprises of a circularly symmetric sunspot with fan-like loop structures visible on one side in the corresponding coronal channels (see Figure~\ref{fig1}). A time-lapse movie of the region, as seen through the AIA~171{\,}{\AA} filter (available online), clearly shows outwardly propagating waves in all of the fan loops. 
\begin{figure}
\centering
\includegraphics[scale=0.7]{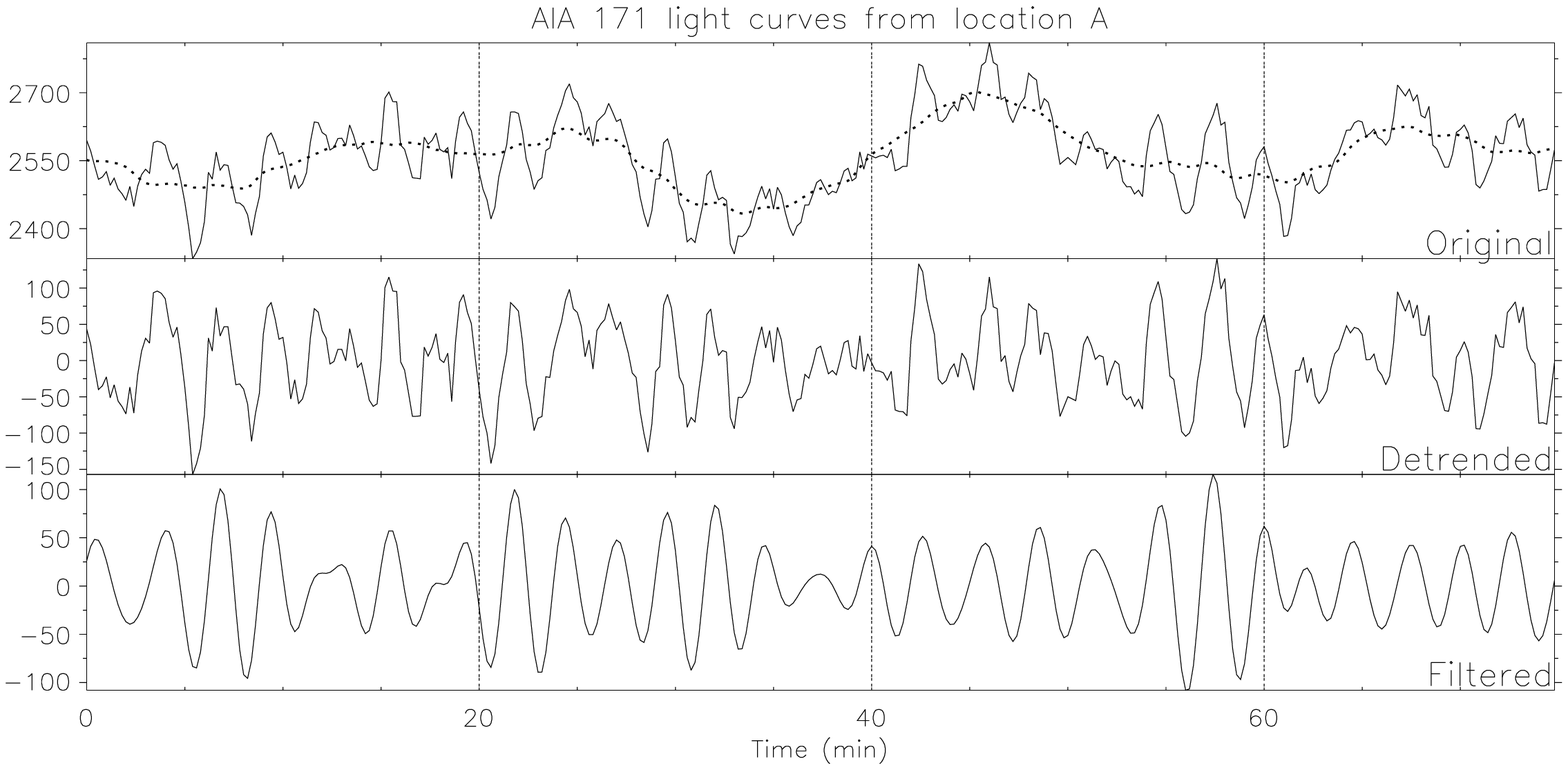}
\caption{Lightcurves from location `A' (see Figure~\ref{fig1}) as captured by the AIA~171{\,}{\AA} channel. The top panel displays the original lightcurve after 3$\times$3 pixel$^2$ binning. The low-frequency background trend is overplotted using a dotted line, which is subsequently subtracted from the original time series to produce the detrended lightcurve shown in the middle panel. The bottom panel displays the reconstructed lightcurve following narrowband Fourier filtration around 3~min.}
\label{fig2}
\end{figure}
Figure~\ref{fig1} indicates a number of preselected regions of interest, labelled as `A', `B' and `C', with the corresponding lightcurves extracted from location `A' detailed in Figure~\ref{fig2}. In this figure, the top panel displays the original lightcurve (after binning over 3$\times$3 pixel$^2$), where an oscillation of $\approx$3~min is visible along with other longer period components. The overplotted dotted line follows the low-frequency background trend, which is subtracted to filter such longer periods, with the resultant displayed in the middle panel. The bottom panel displays the reconstructed lightcurve obtained following Fourier filtration, allowing only a narrow band ($\pm$50~s full width) of frequencies around 3~min. All lightcurves show the persistent presence of 3~min oscillations throughout the duration of the observing sequence. An interesting aspect to note here is the amplitude of the oscillations, which appear to increase and decrease over time. This feature is visible in all the lightcurves, hence ruling out the possibility that these are artefacts of the applied Fourier filters. We also observed this behavior at other locations along the wave propagation path.

\begin{figure}
\centering
\includegraphics[scale=0.48]{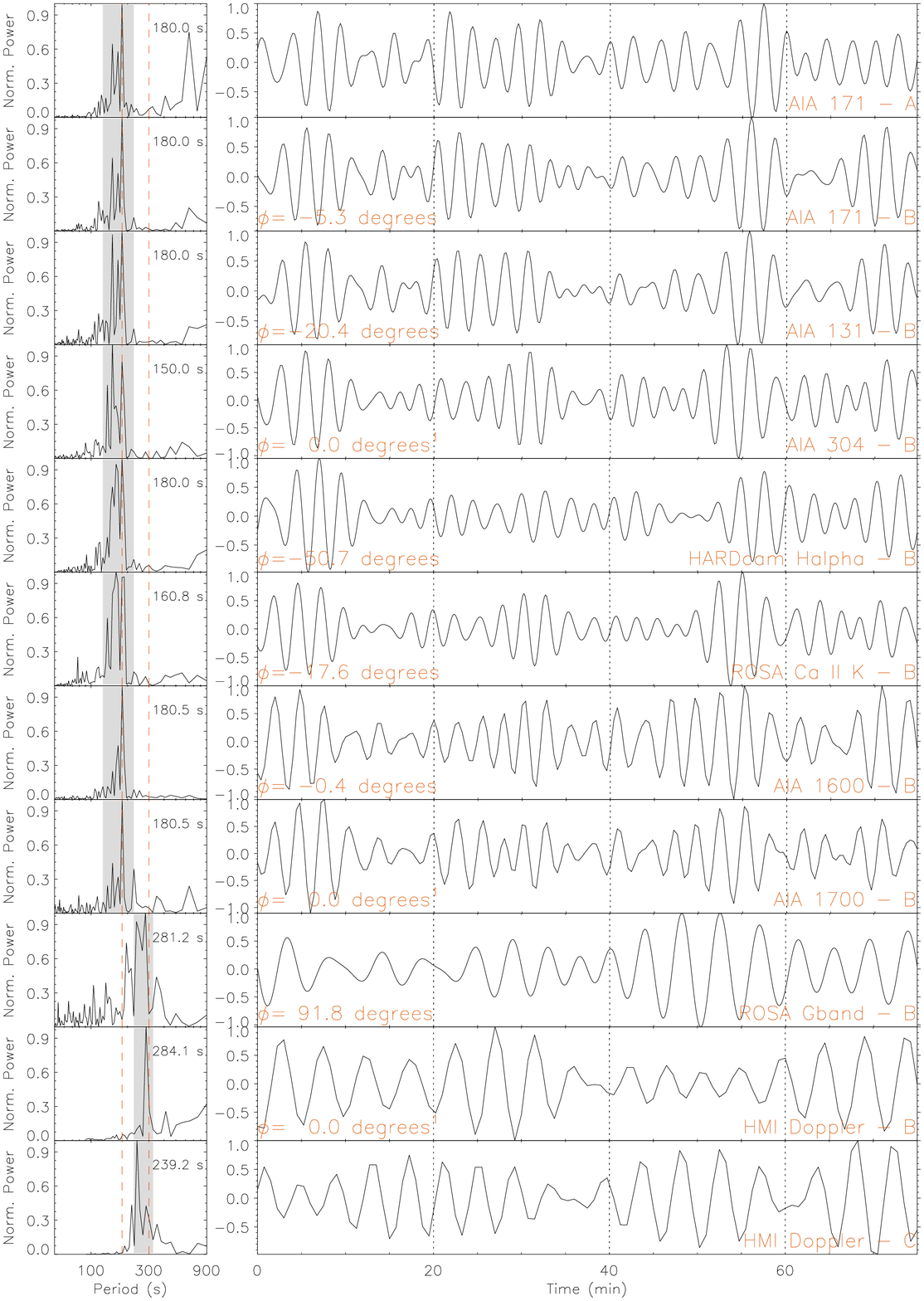}
\caption{\textit{Left}: Normalized Fourier power spectra of original lightcurves from different SDO and ROSA/HARDcam channels. The peak periodicity identified is listed in each plot in seconds. The two vertical dashed red lines mark the locations of 3~min and 5~min periodicities. Regions in grey denote the corresponding frequency band used to produce the filtered light curves shown on right. \textit{Right}: Fourier-filtered lightcurves from the footpoint of the fan loop (location `B' in Figure~\ref{fig1}), plotted as a function of atmospheric height. An AIA~171{\,}{\AA} lightcurve from location `A' and an HMI Dopplergram time series from location `C' are also shown for comparison. The obtained phase angles ($\phi$) for the lightcurves at location B, taking the HMI Dopplergram, AIA 1700{\,}{\AA} and AIA 304{\,}{\AA} channels as a reference (indicated by superscript `1') are also listed.} 
\label{fig3}
\end{figure}

The observed oscillations are similar to those previously studied by many authors \citep[e.g.,][to name but a few]{dihw2002A&A387L13, 2006ApJ...643..540M, 2012ApJ...757..160J, 2012SoPh..281...67K, 2014ApJ...789..118K}, in fan-like loop structures. There has been a debate on whether these oscillations are due to waves or high-speed quasi-periodic upflows \citep{2012RSPTA.370.3193D}, but it is widely believed that the three-minute sunspot oscillations (as presented here) are the signature of propagating slow magneto-acoustic waves. The disappearance of the oscillations after a certain length along the structure further confirms their propagating nature, and with temporal variations in amplitude observed along their full propagation path, such modulation may be a property of the source itself. In fact, similar modulations were found in sub-coronal sunspot oscillations by \citet{1972SoPh...27...61B}, \citet{1982ApJ...253..939G}, \citet{1984ApJ...277..874L}, \citet{1987ApJ...312..457T}, \citet{2001A&A...368..639F}, \citet{2006ApJ...643..540M} and \citet{2006ApJ...640.1153C, 2009ApJ...692.1211C}. Thus, an important aspect is to try and identify the wave source by tracking these oscillations along the loop to its base, and determining what feature(s) and atmospheric height(s) modulate the observed wave trains. It appears that the current fan loop is rooted in the sunspot umbra, which explains the prevalence of 3~min oscillations, and through visual inspection this loop appears to be terminated at the location marked `B' in Figure~\ref{fig1}. Therefore, we constructed similar Fourier-filtered lightcurves from location `B' near the coronal footpoint of the loop (marked in Figure~\ref{fig1}) in all channels representing different layers of the solar atmosphere, with the resulting lightcurves displayed in Figure~\ref{fig3}. Although it is generally assumed that magnetic field lines in the umbral region are mostly vertical, an important aspect to consider is whether location `B' corresponds to the same magnetic feature when observed in all AIA channels. To determine the answer, we used the vector magnetograms from this region and employed nonlinear force-free field extrapolations \citep{2012ApJ...760...47G}, to reveal that the central pixel position of location `B' laterally shifts by less than 3~AIA pixels over a height of 3600~km above the photosphere, thus justifying our choice of a 3$\times$3 pixel$^2$ binning region for all channels. For ROSA and HARDcam channels, we used equivalent macro-pixels to construct the corresponding lightcurves.
The AIA~171{\,}{\AA} lightcurve from location `A' is also shown in this figure, for comparison. Interestingly, the dominant oscillation period at location `B' is $\approx$5~min in the photospheric channels, even within the umbra, which is similar to that reported earlier by several authors \citep{1972SoPh...27...80B, 1976A&A....50..367S, 1981BAAS...13..858T}. The normalized Fourier power spectra generated from the corresponding original lightcurves are shown in left panels of Figure~\ref{fig3}, where the numbers listed in each panel represent the peak periodicity value in seconds. The positions of the 3~min and 5~min periods are marked by vertical dashed red lines for comparison. It is evident that the oscillation power peaks near 3~min for formation heights above that of the AIA~1700{\,}{\AA} and 1600{\,}{\AA} channels, which form near the temperature minimum region, before shifting to periodicities $\approx$5~min in lower atmospheric regions. The ROSA G-band and HMI Dopplergrams form approximately 100~km above photosphere \citep{2012ApJ...746..183J, 2011SoPh..271...27F}. It is noted that $\approx$5~min peaks in G-band imaging and HMI Dopplergrams are accompanied by multiple closely-spaced peaks (similar to those at $\approx$3~min in the other channels), yet there is no significant enhancement at 3~min in these channels. Therefore, to accommodate this frequency shift the reconstructed lightcurves displayed in Figure~\ref{fig3} are Fourier-filtered around 5~min for these two photospheric channels. The frequency band used in each channel to produce the filtered light curves is marked in grey over the corresponding power spectra. Furthermore, to check the nature of propagation of these oscillations from the photosphere to the corona, we estimated phase angles ($\phi$) for the lightcurves at location `B' taking information from the HMI Dopplergram, AIA 1700{\,}{\AA} and AIA 304{\,}{\AA} channels as reference points. The obtained values are listed in Figure~\ref{fig3}. Some channels cannot be used as a reference for photospheric and chromospheric propagation since the periodicities are different. As the observations enter an optically thin regime, the AIA 304{\,}{\AA} channel is chosen as a reference to indicate the coronal propagation of the waves.

From Figure~{\ref{fig3}} it is clear that all lightcurves show a similar modulation in amplitude with time, indicating a possible connection between the 5~min photospheric oscillations and their slow magneto-acoustic counterparts observed in the corona. To extend this further and check the possible connection with photospheric $p$-modes, we identified a location (marked as C in Fig.~\ref{fig1}) outside the sunspot where the average absolute magnetic field strength is $<$20~G, and constructed a similarly Fourier-filtered time series corresponding to that location. This lightcurve is shown in the bottom-right panel of Figure~\ref{fig3}, and interestingly, also displays a similar modulation in its oscillatory amplitude. 
\begin{figure}
\centering
\includegraphics[scale=0.6]{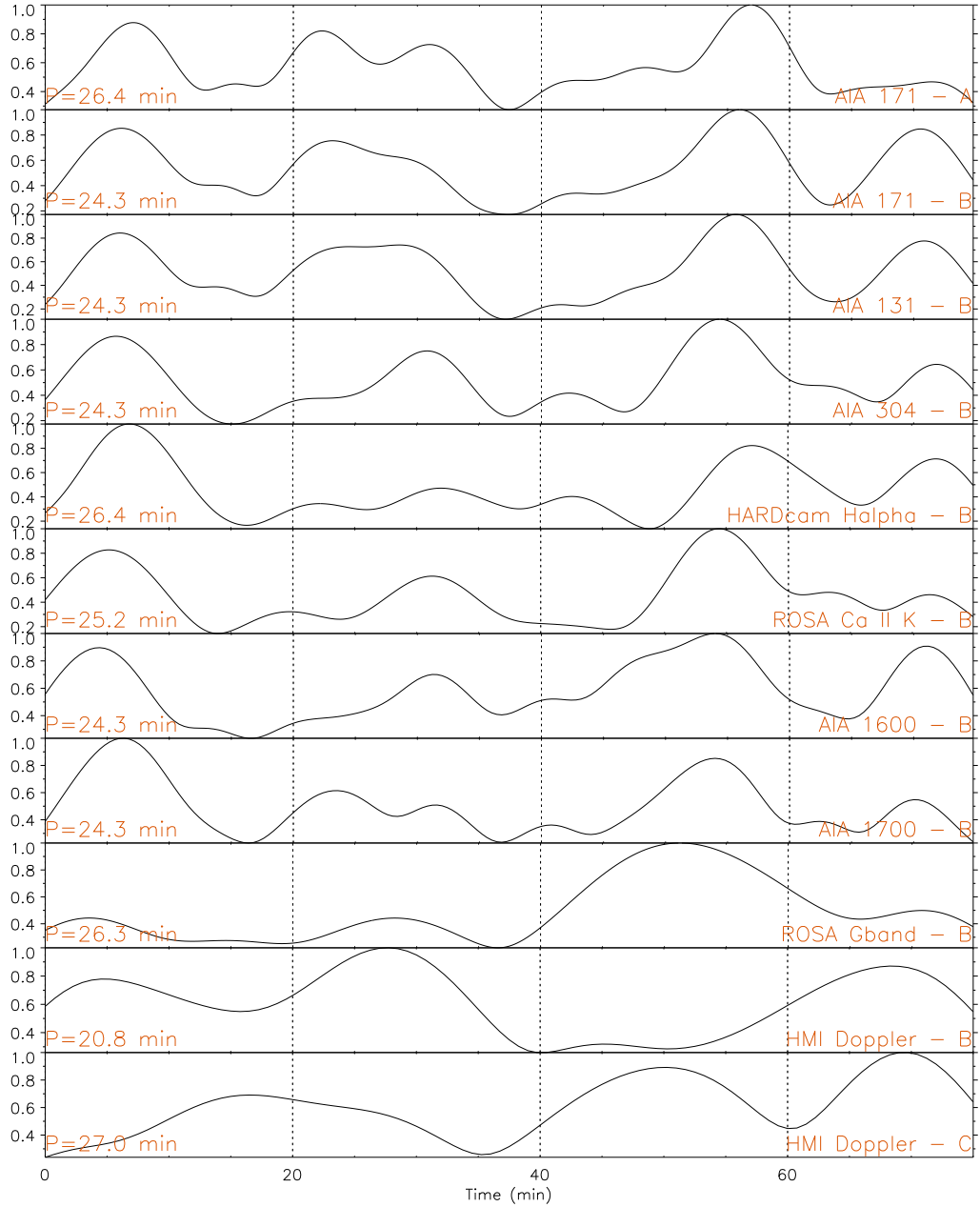}
\caption{Oscillations amplitude variations, extracted from the Fourier-filtered lightcurves shown in Figure~\ref{fig3}, are displayed in ascending atmospheric height order. An estimate of the time scales over which each amplitude modulation occurs is listed for each panel.}
\label{fig4}
\end{figure}
In order to quantify the observed variation in amplitude, we employed wavelet analysis over the filtered lightcurves and extracted the amplitude variation (i.e., correlated with the wavelet power) as a function of time. The obtained amplitudes are shown in Figure~\ref{fig4}. By applying wavelet analysis to the obtained amplitudes, we estimated the dominant period at which the amplitudes vary. This value is  listed at the bottom-left corner of each panel in Figure~\ref{fig4}. The amplitude fluctuations appear to have a reciprocating nature with mean values on the order of $20-27$~min. It may be noted that since the amplitudes are extracted from the narrowband Fourier-filtered lightcurves, the observed fluctuations purely represent the variations in amplitude of 3 (or 5) min oscillations and thus have distinguished themselves from the usually present long period sunspot oscillations (see Figure~\ref{fig2}, top panel). 
\begin{figure}
\centering
\includegraphics[]{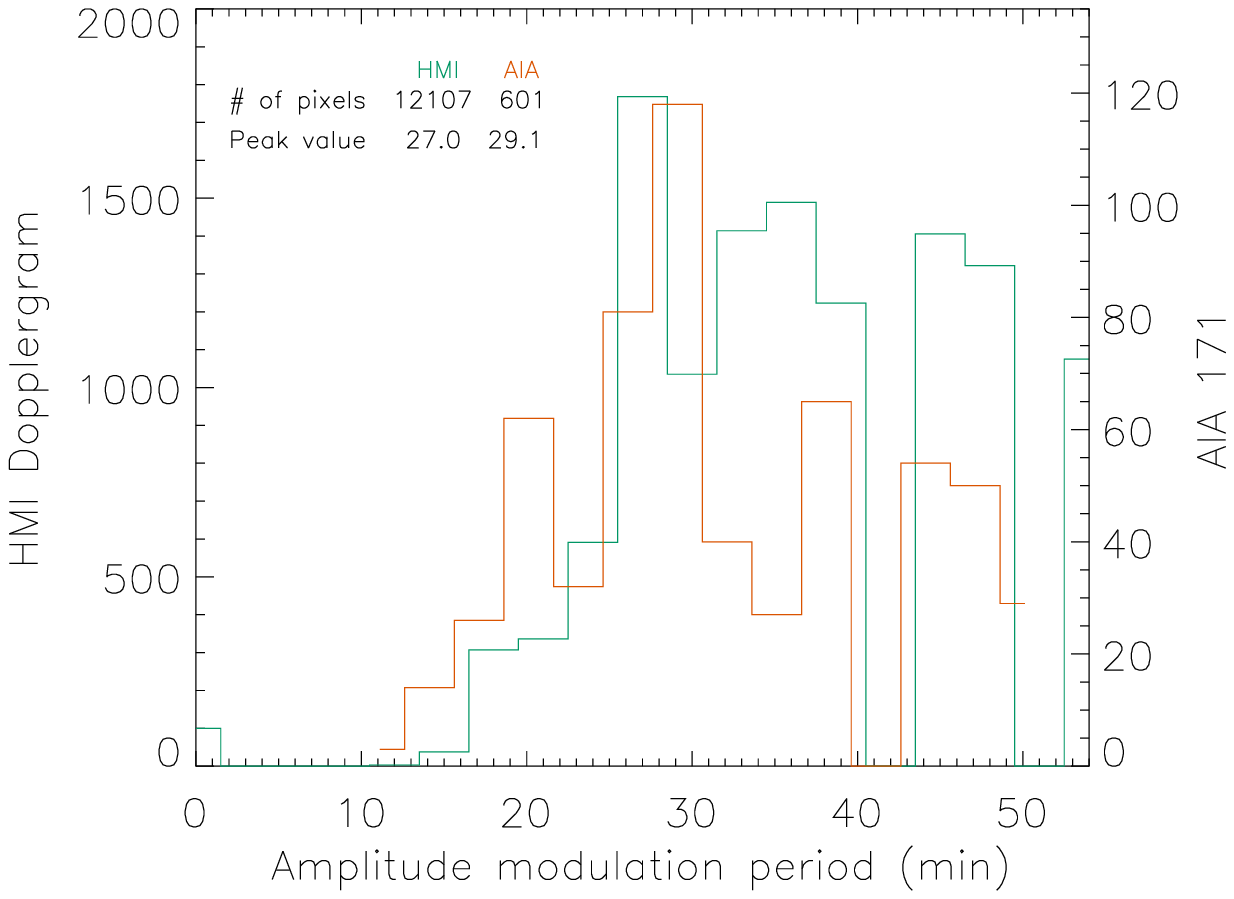}
\caption{Histograms of amplitude modulation period from all the pixels in AIA 171 that display dominant 3~min oscillations and all the pixels in HMI Dopplergram that has magnetic field strength $<$20~G.}
\label{fig5}
\end{figure}
To verify if other loops in the coronal fan system behave in a similar way, we performed identical analysis at all pixel locations in the AIA 171{\,}{\AA} channel that displayed prominent 3~min oscillations ($\sim$600 pixel locations), as well as all pixel locations in the HMI Dopplergrams that contained average absolute magnetic field strengths $<$ 20~G (over 12000 pixel locations), and calculated their corresponding amplitude modulation periods. The results are displayed in a histogram plot shown in Figure~\ref{fig5}, which reveals peak modulation periods of 27.0~min and 29.1~min (modal values) in the HMI and AIA channels, respectively, thus strengthening our hypothesis that the global p-mode oscillations and the propagating waves found in coronal loop systems are related. 
 
\section{Discussion and Conclusions}
Propagating slow magneto-acoustic waves observed in fan-like loop structures are found to exhibit temporal variations in their amplitude. We employed multi-wavelength and multi-instrument data to track the oscillations along the length of the loop structure. Similar modulations in the wave amplitudes are observed at all atmospheric heights, including at the base of the photosphere. The amplitude fluctuations appear to be modulated with a periodicity on the order of $20-27$~min across all bandpasses. Despite the fan loops being anchored in the sunspot umbra, the dominant oscillation period at its base shifts from 3~min to 5~min close to the photosphere, while the amplitude modulation period remains the same. The 5~min Doppler oscillations from a non-magnetic region outside the active region also show similar variations in wave amplitude across the same period range. This behavior is also observed in other parts of the fan loop system, which show a good correlation in amplitude modulation periods with that of non-mangetic regions outside the sunspot in HMI Dopplergrams. These results highlight a possible connection between the photospheric $p$-modes and the propagating slow magneto-acoustic waves observed in the corona.

Amplitude modulation has previously been observed in sunspot oscillations by several authors \citep{1972SoPh...27...61B, 1982ApJ...253..939G, 1984ApJ...277..874L, 1987ApJ...312..457T, 2001A&A...368..639F, 2006ApJ...643..540M, 2006ApJ...640.1153C, 2009ApJ...692.1211C}, and it has been suggested their existence is caused by a number of closely-spaced frequencies leading to a beat phenomenon. The formation of multiple closely-spaced frequencies is explained by resonant filter theories \citep{2005A&A...433.1127Z}, with \citet{2008ApJ...681..672M} employing a Bayesian model to resolve four closely-spaced frequencies in the transition region oscillations that demonstrated amplitude modulations. They identified that the frequencies were part of the global $p$-mode spectrum and suggested a link between the photosphere and the upper atmosphere. Indeed, we find multiple closely-spaced peaks in the power spectra of oscillations detected over all atmospheric heights (left panels of Figure 3), which helps explain the amplitude modulations and reveals the propagation of photospheric $p$-modes up to coronal heights. Furthermore, the negative phase angles obtained in the chromospheric and coronal channels also suggest the upward propagation of waves. The nearly 90$^{\circ}$ phase difference observed between the G-band intensities and the HMI Doppler velocities, which originate at approximately the same atmospheric height ($\approx$100~km), might probably indicate some reflection at the temperature minimum region. 

As suggested by many theoretical models \citep[e.g.,][]{2008ApJ...677..769H, 2014ApJ...796...72J}, the photospheric $p$-modes are likely the source of propagating slow magneto-acoustic waves in sunspots which are guided up to coronal heights along the strong umbral field lines, permeating coronal fans before becoming dissipated. Similarly, those upwardly propagating magneto-acoustic waves that travel along more inclined fields (e.g., such as those towards the umbra/penumbra boundary) end up as running penumbral waves in the chromosphere \citep[e.g.,][]{2006SoPh..238..231K, 2013ApJ...779..168J}. The shift in oscillation period from 5~min to 3~min in chromosphere can either be due to the creation of a chromospheric resonance cavity \citep{1983SoPh...82..369Z} or a result of the natural resonant excitation at the atmospheric cut-off frequency which is higher within the upper photosphere \citep{1991A&A...250..235F} or due to a simple interplay between the maximum power and the cut-off frequency \citep[][]{2015lsrp_tmp}.

\acknowledgements 
The authors thank the referee for useful comments. D.B.J. would like to thank STFC for an Ernest Rutherford Fellowship, in addition to a dedicated standard grant which allowed this project to be undertaken. The AIA and HMI data used here are courtesy of NASA/SDO and the AIA and HMI science teams.

{\it Facilities:} \facility{Dunn (ROSA, HARDcam)}, \facility{SDO (AIA, HMI)}.


\end{document}